\documentclass[11pt]{article}
\usepackage{moriond,epsfig,url}

\def\be{\begin{equation}}
\def\ee{\end{equation}}
\def\bea{\begin{eqnarray}}
\def\eea{\end{eqnarray}}

\newcommand{\eg}{e.g.\ }
\newcommand{\cG}{{\cal G}}
\newcommand{\order}[1]{\mathcal{O}\left(#1\right)}
\begin{document}
\begin{flushright}
LPTHE-P06-04
\end{flushright}

\vspace*{2.5cm}
\title{FASTJET: A CODE FOR FAST $k_t$ CLUSTERING, AND MORE~\footnote{Talk give at
Moriond QCD, La Thuile (Italy), March 2006, and DIS2006, Tsukuba (Japan) April
2006}}

\author{ MATTEO CACCIARI}

\address{LPTHE, Universit\'e P. et M. Curie - Paris 6, France}

\maketitle\abstracts{
Two main classes of jet clustering algorithms, cone and $k_t$, 
are briefly discussed. It is argued that the former can be often cumbersome to
define and implement, and difficult to analyze in terms of its behaviour with
respect to soft and collinear emissions. The latter, on the other hand, enjoys a
very simple definition, and can be easily shown to be infrared and collinear
safe.  Its single potential shortcoming, a computational complexity 
believed  to scale like the number of particles to the cube ($N^3$), is 
overcome by  introducing a new geometrical algorithm that reduces it to 
$N\ln N$. A 
practical implementation of this approach to $k_t$-clustering, {\tt FastJet}, 
is shown to be
orders of magnitude faster than all other present codes, opening the
way to the use of $k_t$-clustering even in highly populated heavy ion events.}


High energy events are often studied in terms of jets. While a ``jet'' is in
principle just a roughly collimated bunch of particles flying 
in the same direction, it takes of course a more careful definition
to make it a tool for an accurate analysis of QCD. In particular, in order
to be able to compare the experimentally observed jets to theoretical
predictions, one must ensure that the measured quantity is ``soft and collinear
safe'', meaning that the addition of a soft or a collinear parton does not
change its value. Only for this type of quantity can higher order calculations in
QCD  give sensible results.

While jets have been discussed since the beginning of the '70s, the first
modern definition of a soft and collinear safe jet is due to Sterman and
Weinberg~\cite{Sterman:1977wj}.
Their jets, whose definition was originally formulated for $e^+e^-$
collisions, were of a kind which became successively known as
`cone-type'.  They have been successively extended to hadronic collisions, 
where cone-type jets are based on identifying energy-flow into cones in 
(pseudo)rapidity
$\eta=-\ln \tan \theta/2$ and azimuth $\phi$, together with various steps of
iteration, merging and splitting of the cones to obtain the final
jets.  The freedom in the details of the clustering procedure  
has led to a number of definitions of cone-type jet clustering 
algorithms, many of them
currently used at the Tevatron and in preliminary studies of LHC
analyses~\cite{Cone}.
However, cone jet-finders tend to be rather complex: 
different
experiments have used different variants (some of them infrared unsafe),
and it is often difficult to know exactly which jet-finder to use in
theoretical comparisons. 

Partly in order to overcome these difficulties,
at the beginning of the '90s cluster-type jet-finders where proposed. They are 
generally based on successive
pair-wise recombination of particles, 
have simple definitions and are all infrared safe 
. The most widely used of them is the 
$k_t$ jet-finder~\cite{Kt}, defined below. Among its physics
advantages are  (a) that it
purposely mimics a walk backwards through the QCD branching sequence, 
which means that reconstructed jets naturally collect most of the
particles radiated from an original hard parton, giving better
particle mass measurements~\cite{KtVersusCone,KtVersusConeBis} and
gaps-between-jets identification~\cite{KtGaps} (of relevance
to Higgs searches); 
and (b) it allows one
to decompose a jet into constituent subjets, which is useful for
identifying decay products of fast-moving heavy particles (see 
e.g.~\cite{KtSubJetAnalysis}) and various QCD studies. This has led to the
widespread adoption of the $k_t$ jet-finder in the LEP ($e^+e^-$
collisions) and HERA ($ep$) communities.

The $k_t$ jet-finder, in the longitudinally invariant formulation
suitable for hadron colliders, is defined as follows. 

\noindent
\begin{center}
\begin{tabular*}{0.9\textwidth}{c}
\hline
The $k_t$ jet-finder\\
\hline
\begin{minipage}{0.8\textwidth}
\begin{itemize}
\item[1.] For each pair of particles $i$, $j$ work out the $k_t$
  distance $d_{ij} = \min(k_{ti}^2,{k_{tj}^2}) R_{ij}^2$ with
  $R_{ij}^2 = (\eta_i-\eta_j)^2 + (\phi_i-\phi_j)^2$, where $k_{ti}$,
  $\eta_i$ and $\phi_i$ are the transverse momentum, rapidity and
  azimuth of particle $i$; for each parton $i$ also work out the beam
  distance $d_{iB} = k_{ti}^2$.
\item[2.] Find the minimum $d_{\min}$ of all the $d_{ij},d_{iB}$. If
  $d_{\min}$ is a $d_{ij}$ merge particles $i$ and $j$ into a single
  particle, summing their four-momenta (alternative recombination
  schemes are possible); if it is a $d_{iB}$ then declare particle $i$
  to be a final jet and remove it from the list.
\item[3.] Repeat from step 1 until no particles are left.
\end{itemize}
\end{minipage}
\\
\hline
\end{tabular*}
\vspace{10pt}
\end{center}

One apparent drawback of this algorithm is its computational
complexity, originally believed to scale like $N^3$, $N$ being the
number of particles to be clustered. This complexity leads to concrete
implementations which become slow as $N$ grows, making the use of
$k_t$-clustering impractical in environments where large numbers of
particles are produced in the final state, like hadron-hadron or, even
more spectacularly, ion-ion collisions.

We show here that this computational complexity can in fact be reduced
to $N\ln N$, opening the way to a much more widespread use of the $k_t$
jet-finder~\cite{Cacciari:2005hq}.

To obtain a better algorithm we isolate the geometrical aspects of the
problem, with the help of the following observation (see~\cite{Cacciari:2005hq}
for its proof): If $i$, $j$ form the smallest $d_{ij}$, and $k_{ti} <
k_{tj}$, then $R_{ij} < R_{i\ell}$ for all $\ell \neq j$, i.e. $j$ is
the geometrical nearest neighbour of particle $i$.

This means that if we can identify each particle's
geometrical nearest
neighbour (in terms of the geometrical $R_{ij}$ distance), then we
need not construct a size-$N^2$ table of 
$d_{ij} = \min(k_{ti}^2,{k_{tj}^2}) R_{ij}^2$, but only the
size-$N$ array, $d_{i{\cG\!}_i}$, where $\cG_i$ is $i$'s $\cG$eometrical 
nearest neighbour\footnote{We shall drop `geometrical' in the
following, speaking simply of a `nearest neighbour'}.
We can therefore write the following algorithm~\cite{Cacciari:2005hq}:
\noindent
\begin{center}
\begin{tabular*}{0.9\textwidth}{c}
\hline
The {\tt FastJet} Algorithm\\
\hline
\begin{minipage}{0.8\textwidth}
\begin{enumerate}
\item For each particle $i$ establish its nearest neighbour $\cG_i$ and
  construct the arrays of the $d_{i{\cG\!}_i}$ and $d_{iB}$.
\item Find the minimal value $d_{\min}$ of the $d_{i{\cG\!}_i}$, $d_{iB}$.
\item Merge or remove the particles corresponding to $d_{\min}$ as
  appropriate.
\item Identify which particles' nearest neighbours have changed and
  update the arrays of $d_{i{\cG\!}_i}$ and $d_{iB}$. If any particles are
  left go to step 2.
\end{enumerate}
\end{minipage}
\\
\hline
\end{tabular*}
\vspace{10pt}
\end{center}

This already reduces the problem to one of
complexity $N^2$: for each particle we can find its  
nearest neighbour
by scanning through all $\order{N}$ other particles [$\order{N^2}$
operations]; calculating the $d_{i{\cG\!}_i}$, $d_{iB}$ requires $\order{N}$
operations; scanning through the $d_{i{\cG\!}_i}$, $d_{iB}$ to find the
minimal value $d_{\min}$ takes $\order{N}$ operations [to be repeated
$N$ times]; and after a merging or removal, updating the nearest
neighbour information will require $\order{N}$ operations [to be
repeated $N$ times].

\begin{figure}[t]
\begin{minipage}{7cm}
\vspace{-6cm}
\epsfig{file=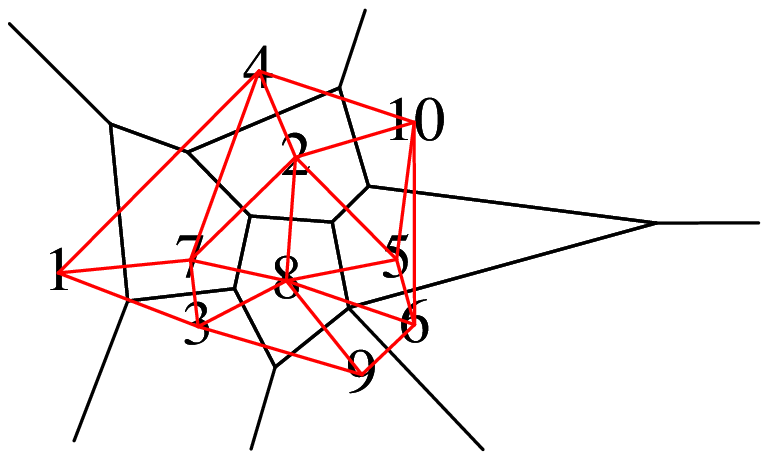,width=6.5cm}
\end{minipage}
~~~~~
\epsfig{file=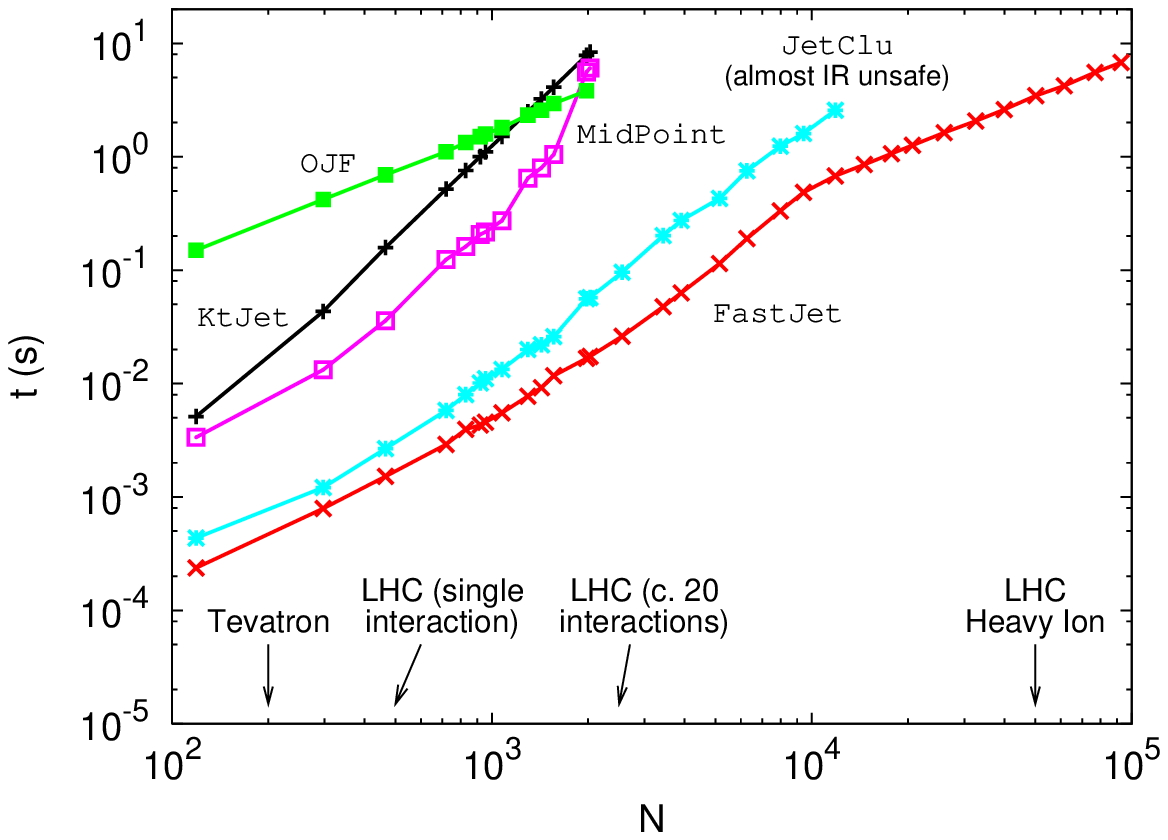,width=7cm}
\vspace{-.3cm}
\caption{\label{fig1} Left: the Voronoi diagram (black lines) 
of ten points in a plane, numbered
1...10. Superimposed, in red, is the Delaunay triangulation. Right: CPU
time taken to cluster $N$ particles for various jet-finders. {\tt FastJet}
 is
available at \protect\url{http://www.lpthe.jussieu.fr/~salam/fastjet}.
}
\end{figure}

We note, though, that three steps of this algorithm ---
initial nearest neighbour identification, finding $d_{\min}$ at each
iteration, and updating the nearest neighbour information at each
iteration --- bear close resemblance to problems
studied in the computer science literature and for which efficient
solutions are known. An example is the use of a structure known as a Voronoi
diagram~\cite{WikipediaRefs} or its dual, a Delaunay triangulation (see
fig.~\ref{fig1}), to
find the nearest neighbour of each element of an ensemble of vertices in
a plane (specified by the $\eta_i$ and
  $\phi_i$ of the particles). It can be shown that such a structure can
  be built with $\order{N
    \ln N}$ operations (see \eg~\cite{Fortune}), and updated with 
  $\order{\ln N}$ operations~\cite{DelaunayDeletion}
  (to be repeated $N$ times).
  More details, concerning also other steps in the algorithm, are given 
  in~\cite{Cacciari:2005hq}. The final result is
  that  both the geometrical and minimum-finding aspects of the
$k_t$ jet-finder can be related to known problems whose solutions
require $\order{N \ln N}$ operations.

The FastJet algorithm has been implemented in the  C++ code {\tt
FastJet}. The building and the updating of the Voronoi diagram have been
performed using the publicly available  Computational
Geometry Algorithms Library (CGAL)~\cite{CGAL}, in particular its
triangulation components~\cite{CGALTriang}.
The resulting running time for the clustering of $N$ particles is displayed in
fig.~\ref{fig1}. It can be seen to be faster than all other codes
currently used, both of cone or $k_t$ type. Analyses of events with
extremely high multiplicity, like heavy ion collisions at the LHC, are
now feasible, their clustering taking only about 1 second, rather than 1
day of CPU time.

\begin{center}
\begin{figure}[t]
\epsfig{file=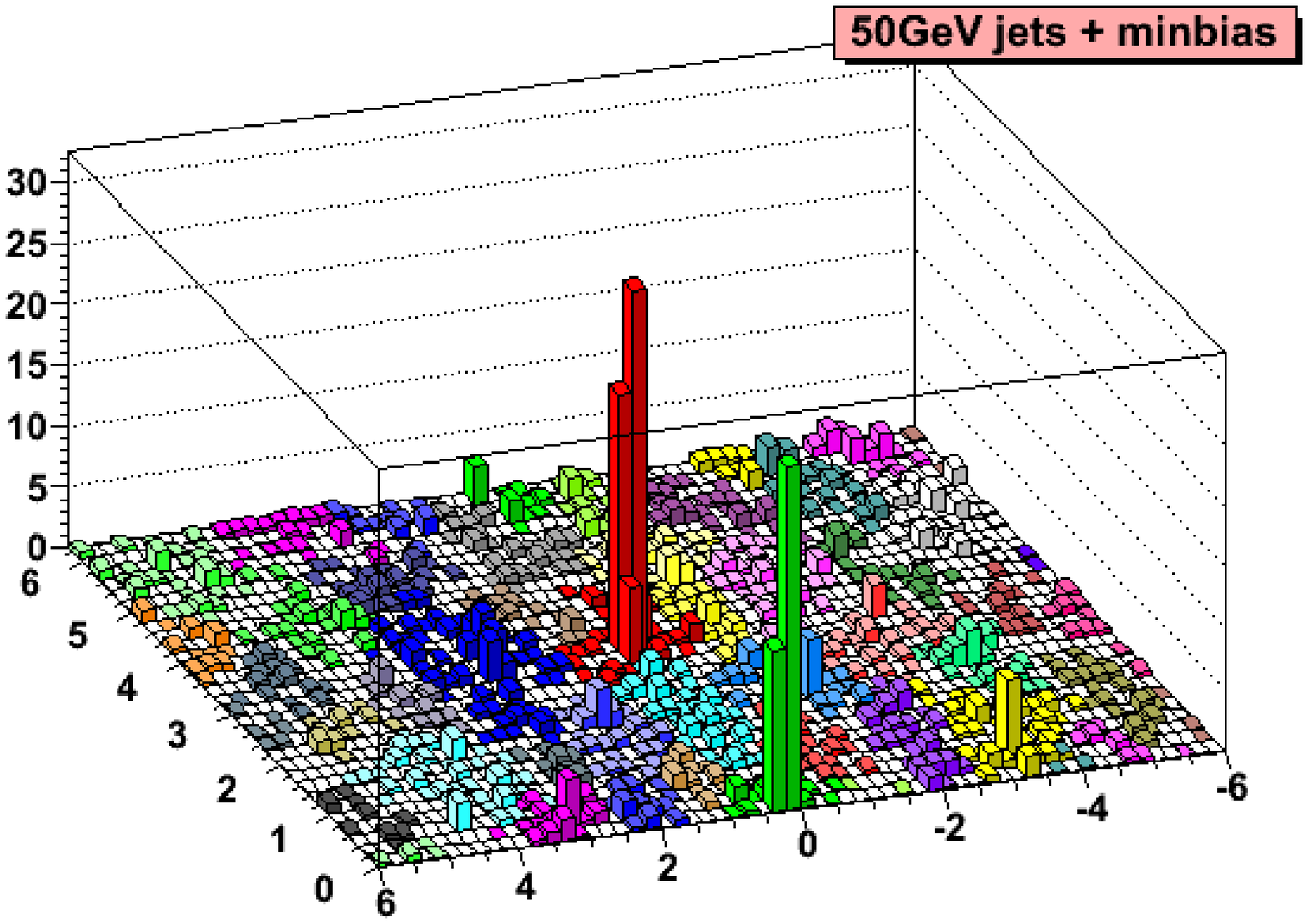,width=7.5cm}~~
\epsfig{file=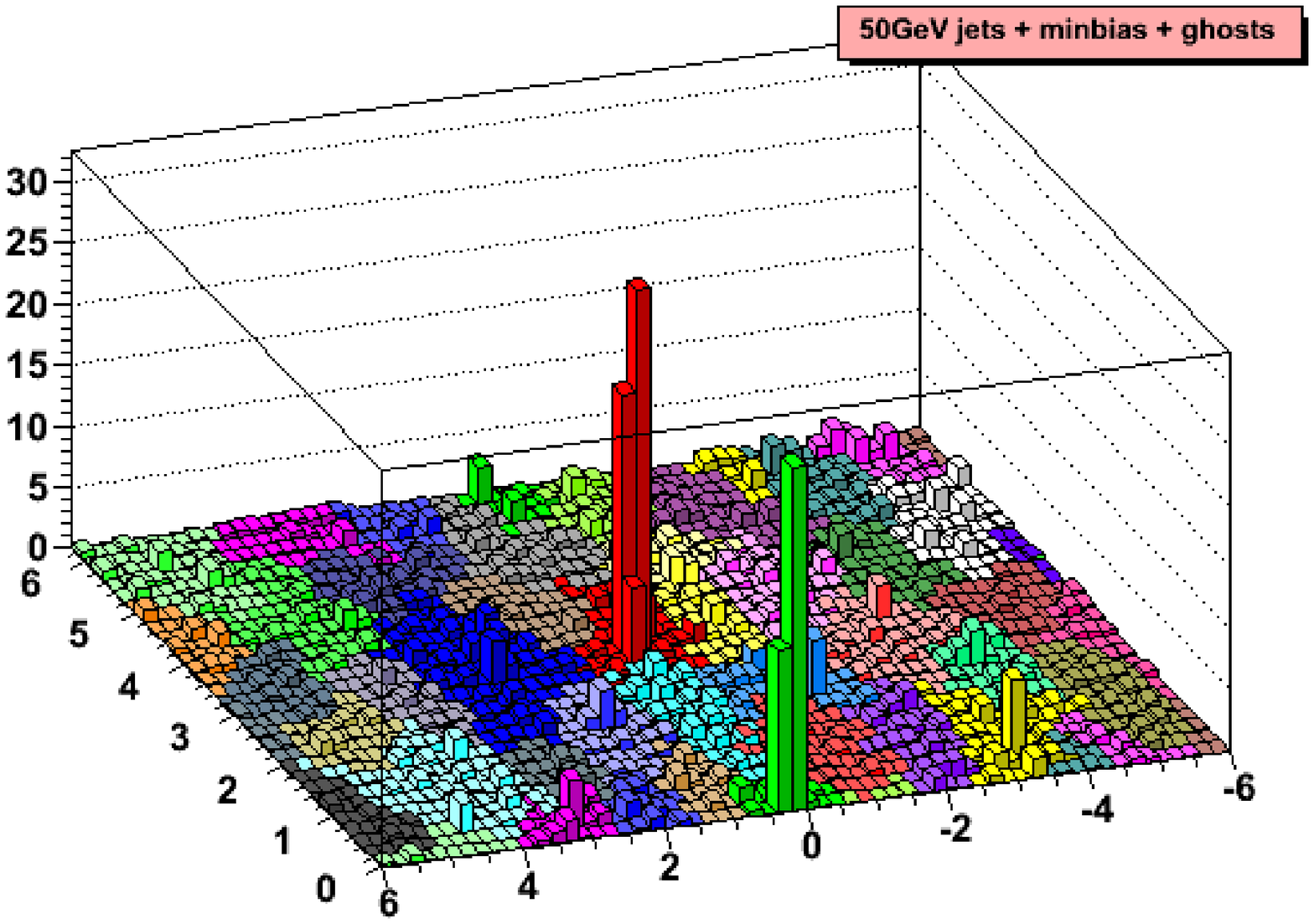,width=8cm}
\vspace{-.7cm}
\caption{\label{fig2} 
A simulated ``typical'' event at high luminosity at the LHC. 
Left: A single event with two hard jets has been combined with about 10 softer
events. Right: Very soft `ghost' particles have  been added in order to be
able to quantify more precisely the area of each jet.}
\end{figure}
\end{center}

The speed of  {\tt FastJet}  does more, however, than just making analyses
with a few hundred particles faster, or  those with a few thousand
possible. In
fact, it allows one to do {\sl new things}. 
One example is the possibility of calculating the {\sl area}
of each jet by adding to the event a large number of extremely soft
`ghost' particles, and counting how many get clustered into any given jet.
This approach is of course computationally heavy, and would  be
unfeasible -- or at least extremely impractical --  with a slower jet-finder.
Fig.~\ref{fig2} shows the result of this procedure on a LHC event made
of one hard and many soft jets. Estimating jet areas is of
course not interesting by itself, but as an intermediate step towards
performing an event-by-event subtraction of underlying event/minimum
bias energy from the hard jets.  This work is presently in
progress~\cite{cacciarisalam2}.

\vspace{-10pt}
\noindent
{\bf Acknowledgements.} I wish to thank Gavin Salam for the ongoing entertaining
collaboration on this project.

\end{document}